\documentclass[]{spie}  
\usepackage[]{graphicx}
\usepackage{latexsym}

\title{Squeezed Light and Entangled Images from Four-Wave-Mixing in Hot Rubidium Vapor}

\author{Raphael C Pooser, Vincent Boyer, Alberto M. Marino, Paul D.
Lett
 \skiplinehalf Joint Quantum Institute, National Institute of
Standards and Technology, University of Maryland, Gaithersburg, MD
20899, USA}

\authorinfo{Send correspondence to raphael.pooser@nist.gov, paul.lett@nist.gov}

\begin{document}
  \maketitle

\begin{abstract}
Entangled multi-spatial-mode fields have interesting applications in
quantum information, such as parallel quantum information protocols,
quantum computing, and quantum imaging. We study the use of a
nondegenerate four-wave mixing process in rubidium vapor at 795 nm
to demonstrate generation of quantum-entangled images. Owing to the
lack of an optical resonator cavity, the four-wave mixing scheme
generates inherently multi-spatial-mode output fields. We have
verified the presence of entanglement between the multi-mode beams
by analyzing the amplitude difference and the phase sum noise using
a dual homodyne detection scheme, measuring more than 4 dB of
squeezing in both cases. This paper will discuss the quantum
properties of amplifiers based on four-wave-mixing, along with the
multi mode properties of such devices.
\end{abstract}



\section{INTRODUCTION}
\label{sec:intro}  

The quantum correlations and entanglement present in
multi-spatial-mode squeezed fields, which have applications in
quantum information and quantum computing, have recently garnered
increased general interest. Quantum entanglement in the context of
continuous light fields is normally characterized by the fields'
quantum fluctuations. Specifically, the quantum fluctuations of a
pair of entangled modes (i.e. two-mode squeezed states or twin
beams) are correlated to a greater extent than would be possible
when observing classical fluctuations. Such states are interesting
because of their applicability to quantum information protocols
\cite{Braunstein_vanLoock}.
For multi-spatial-mode twin-beams each beam consists of multiple
independent spatial modes. Bipartite entanglement can exist between
the independent spatial modes that make up the overall beams, so
that many spatial mode pairs are entangled together in parallel.
This can give rise to parallel quantum information processing
protocols or even squeezing distillation
\cite{schnabel,leuchsdistill}.

Further, multi-spatial-mode beams of light can carry images; storing
information in their spatial degrees of freedom. Classically,
imaging is thought of as the storage of information in the local
mean intensity of a beam. Holographic imaging stores information in
the classical phase of the light beam as well. However, one can also
consider the local quantum fluctuations of such imaging fields,
leading to the concept of quantum imaging \cite{kobolov}. Quantum
imaging is most generally defined as the storage of information in
the quantum fluctuations of the phase and amplitude of multiple
spatial modes. Accessing the quantum fluctuations of the light
yields a higher information density than would be available in a
classical image, essentially due to an increased signal to noise
ratio stemming from quantum correlations \cite{kobolov_fabre}.
Recently, quantum imaging enabled the detection of transverse beam
displacements smaller than the standard quantum limit
(SQL)\cite{treps}, and noiseless image amplification has also been
demonstrated \cite{mosset_deveaux}. The prospect of generating these
types of light beams is one of the more exciting applications of
multi-spatial-mode amplifiers.

This paper will show the generation of quantum images using a
nonlinear interaction, four-wave-mixing in rubidium vapor. The next
section describes the basic amplifier and its components. The third
section characterizes the multi-spatial-mode aspect of the
amplifier, and the last section puts this characteristic to use in
the generation of entangled images.

\section{Four Wave Mixing Amplifier}
\label{sec:4wm}
\begin{figure}[ht]
\centering
\includegraphics[width=4in]{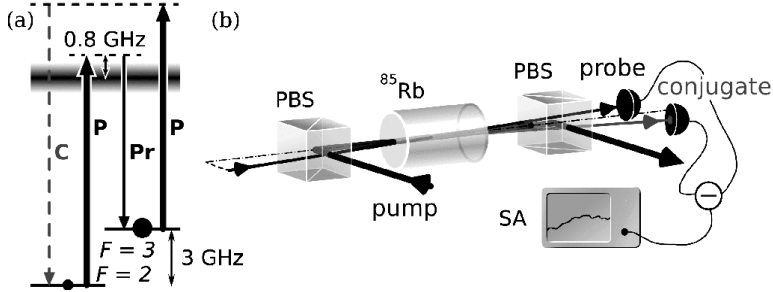}
\caption{ 4WM configuration. (a) Four-level double-lambda scheme in
$^{85}$Rb, P = pump, C = conjugate, Pr = probe.  The width of the
excited state represents the Doppler broadened profile. (b)
Experimental setup, PBS = polarizing beam splitter, SA = spectrum
analyzer. Reproduced from ref.~\citenum{McCormicklowfreq}.
\label{fig:4WMscheme}}
\end{figure}

Four-wave-mixing (4WM) is a nonlinear process based on a third order nonlinearity.
Our 4WM scheme is based on a double-lambda system configuration in $^{85}$Rb, shown in figure \ref{fig:4WMscheme}.
A strong pump field, detuned 800 MHz from the D1
($5S_{1/2},F=2\rightarrow 5P_{1/2}$) transition in $^{85}$Rb (795
nm), and a weak probe beam, detuned 3 GHz to the red of the pump (by
the hyperfine ground-state splitting), are input into the vapor
cell. The strong third order nonlinearity arises from the coherence
built-up between the two hyperfine levels in the ground state. The
amplified probe and generated conjugate beams make up the output
light fields and have been predicted to exhibit quantum noise
reduction \cite{Lukin2000}.

\subsection{Theoretical Two-Mode Quantum Correlations} \label{quantcorr}
In general 4WM four electric fields can interact in configurations
in which one or more of the fields experience amplification or
deamplification inside the nonlinear medium proportional to the
magnitude of the nonlinear susceptibility and the amplitude of the
interacting fields (in the absence of any other interactions between
the light and matter). The nonlinear propagation equations are
\cite{boydbook}
\begin{equation}
\frac{\partial E_i}{\partial t} =  i \kappa_i E_j^*E_kE_l
\label{eq:NLfinal}
\end{equation}
(neglecting other sources of loss and assuming perfect
phase-matching), where $\kappa_i$ is proportional to the third order
nonlinear susceptibility.

Similar to systems that have large second order nonlinearities, one
can envision configurations based on 4WM that result in quantum
correlations between some of the fields. The Heisenberg equations of
motion for the field operators are found by quantizing classical
nonlinear equations:
\begin{equation}
\dot{a_i} =  \chi a_j^\dagger a_k a_l. \label{eq:NLquant}
\end{equation}
Typically 4WM schemes involve large field amplitudes input to the nonlinear medium to be used as a
pump for the nonlinear interaction (as is true for our configuration).
In this situation the pump fields can be considered classical amplitudes,
since the gain or loss in the pump fields will be very small relative to the starting
amplitudes. In eq.\ (\ref{eq:NLquant}) above, if one had two of the fields input as pumps to the
nonlinear process, one would be left with:
\begin{eqnarray}
\dot{a_1} =  \kappa a_2^\dagger, \label{eq:4WMsimple1} \\
\dot{a_2} =  \kappa a_1^\dagger, \label{eq:4WMsimple2}
\end{eqnarray}
where subscripts 1 and 2 represent two separate amplified fields,
for the case in which both pumps are derived from the same electric
field. Eqs.\ (\ref{eq:4WMsimple1}) and (\ref{eq:4WMsimple2}) lead to
quantum correlations between continuous field variables,
specifically linear combinations of the field quadratures. The
quadratures themselves are linear combinations of the raising and
lowering operators for the electric fields. One can define a basis
with two orthogonal field quadratures:
\begin{eqnarray}
X_i=\frac{1}{\sqrt{2}}(\hat{a}_i+\hat{a}_i^\dagger) \label{eq:quadratureX} \\
P_i=\frac{i}{\sqrt{2}}(\hat{a}_i^\dagger - \hat{a}_i),
\label{eq:quadratureP}
\end{eqnarray}
where $i=1,2$ corresponds to the two fields amplified by the
nonlinear process. Substituting into eqs.\ (\ref{eq:4WMsimple1}) and
(\ref{eq:4WMsimple2}) and solving for the quadratures yields
\begin{equation}
P_+(t) = e^{-\kappa t} P_+(0),
\end{equation}
\begin{equation}
X_-(t) = e^{-\kappa t} X_-(0), \label{eq:ampsqueeze}
\end{equation}
where $X_-(t)=X_1(t)-X_2(t)$ and $P_+(t)=P_1(t)+P_2(t)$ are called the generalized quadratures.

The variances of the generalized quadratures decrease as a function of the
interaction time in the nonlinear medium:
$$
V[X_-(t)]=V[X_-(0)]e^{-2\kappa t}.
$$
The variances at $t=0$ are those of the inputs to the nonlinear
medium, typically either the vacuum or a coherent state of nonzero
amplitude. Such inputs are minimum uncertainty states whose noise
levels define the SQL. The variances exhibit the familiar quantum
noise-reduction below the SQL, otherwise called ``squeezing''. Such
noise reduction is often associated with nonlinear interactions
based on parametric down conversion (PDC, second order nonlinearity)
in either optical parametric amplifiers \cite{Heidmann1987} or
optical parametric oscillators \cite{Feng04}. The output states are
the familiar two-mode squeezed states (so-called because the noise
reduction is a property of a joint variable of the two fields).
In the case of 4WM, for every two pump photons that are annihilated
in the nonlinear medium, a single photon is emitted into each of the
two output fields simultaneously. This leads to correlations between
the intensities of the output beams to better than would be allowed
by the SQL.
In other words, one should observe squeezing on the intensity
difference between the output fields.

In fact, squeezing was first observed in 4WM by Slusher  et al.
\cite{Slusher1985}. Since then, noise reduction in 4WM has been
observed under various conditions
\cite{Maeda1987,Orozco1987,Raizen1987,Vallet1990,Ho1991,Hope1992,Josse2003,Ries2003,Hsu2006,Lambrecht1996}.
However, the amount of squeezing had not exceeded 2.2 dB
\cite{Lambrecht1996} until recently \cite{McCormick2007}. Our scheme
has output states which exhibit up to 8 dB of intensity-difference
noise reduction, shown in the next section.

\subsection{Intensity-Difference Noise Reduction}
Light amplified by the 4WM scheme shown in figure
\ref{fig:4WMscheme} can be described by equations
(\ref{eq:4WMsimple1}) and (\ref{eq:4WMsimple2}). A strong pump beam
(400 mW) with a 500 $\mu$m waist and a weak probe (100 $\mu$W) with
a 250 $\mu$m waist are input at a small angle
($\theta\sim0.75^\circ$) so that they overlap throughout the 12.7 mm
length of the vapor cell. The cell is heated to 110 $^\circ$C to
increase the Rb number density (vapor pressure). The probe field is
generated by splitting the pump and passing part of it through an
acousto-optic modulator to downshift its frequency by 3 GHz. As a
result the phase difference between the pump and probe is stable.

In this configuration, the probe field experiences a gain, $G$, and
is amplified, while the conjugate field is amplified from the vacuum
(see fig.\ \ref{fig:4WMscheme}a). The field operators for the probe
and conjugate transform as follows in the gain region:
\begin{eqnarray}
    a_1 \rightarrow a_1\sqrt{G} - a_2^\dagger\sqrt{G - 1} \label{eq:gain1}\\
    a_2^\dagger \rightarrow a_2^\dagger \sqrt{G}-a_1\sqrt{G - 1}
    \label{eq:gain2},
\end{eqnarray}
where $G$ depends on the strength of the interaction:
$\sqrt{G}=\cosh{\kappa t}$, $\sqrt{G-1}=\sinh{\kappa t}$. When the
probe port $a_1$ is seeded with a coherent state $|\alpha\rangle$
and the conjugate port $a_2$ is fed with the vacuum, the output
intensity-difference noise is equal to the shot noise of the input,
which gives a quantum noise reduction of $1/(2G - 1)$ with respect
to the output SQL. Thus, we expect that for a perfect amplifier the
squeezing increases with increasing gain. In reality, losses in the
atomic medium limit the squeezing somewhat, as the probe experiences
more absorption than the conjugate since it is tuned closer to
resonance. A distributed gain/loss model (discussed
elsewhere\cite{McCormicklowfreq}) accurately predicts the amount of
squeezing we measure in the output of our amplifier.

As shown in figure \ref{fig:4WMscheme}b, in our experiment the pump
is separated from the probe and conjugate signal fields on a
polarizing beam splitter, and the probe and conjugate intensities
are detected separately and then subtracted. The difference signal
shows the noise power of the intensity-difference when viewed on a
spectrum analyzer. Figure \ref{fig:int_squeezing} below shows that 8
dB of noise reduction below the SQL is possible with our amplifier
system\cite{McCormicklowfreq}. The actual squeezing bandwidth is
approximately 20 MHz (not shown in fig.\ \ref{fig:int_squeezing}).
\begin{figure}[ht]
    \begin{center}
    \includegraphics[width=4in]{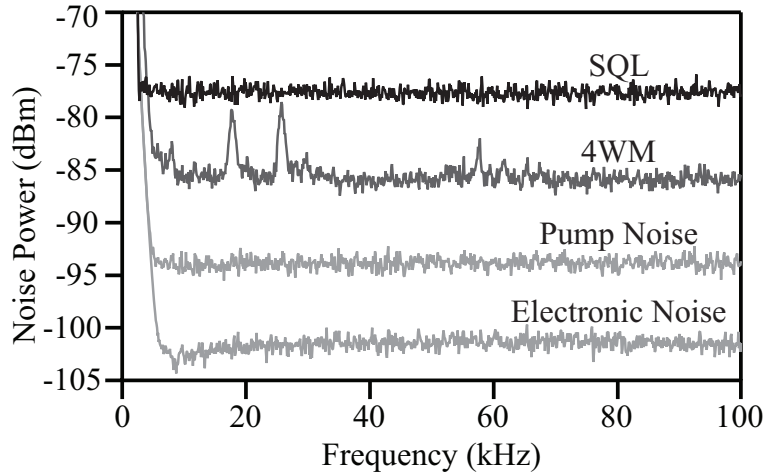}
    \vspace{0.25in}
    \caption{Intensity-difference squeezing. Noise spectra
    (RBW=1~kHz, VBW=10~Hz) for the electronic noise, the scattered pump light,
    the intensity-difference, and the standard-quantum
    limit. Reproduced from ref.~\citenum{McCormicklowfreq}.\label{fig:int_squeezing}}
    \end{center}
\end{figure}

Section \ref{quantcorr} suggests that in addition to the intensities
being correlated, the orthogonal quadratures of the two fields
should also exhibit the same degree of quantum correlations, a
property of phase-insensitive amplifiers\cite{Reid1988}. Further,
the presence of such correlations in both the $X_-$ and $P_+$
quadratures (which can be associated with the intensity-difference
and phase-sum) indicates the presence of entanglement between the
output fields \cite{Reid1988}.


\subsection{Continuous Variable Entanglement} \label{entanglement}
In order to observe entanglement the noise reduction in the two
generalized quadratures must satisfy the inseparability criterion:
$I=\langle \Delta X_-^2 \rangle+\langle \Delta P_+^2 \rangle<2$
\cite{Duan,Simon}. Several other systems have been used to
demonstrate inseparability: optical parametric oscillators below
\cite{Peng1992,Fabre2004} and above \cite{villar,olivier2006}
threshold, 4WM in optical fibers \cite{Silberhorn}, and mixing of
single-mode squeezed states \cite{Furusawa}.

To demonstrate the entanglement between the probe and conjugate
fields of our amplifier, the quadratures of both fields need to be
detected simultaneously. Both signal field quadratures are detected
using homodyne detectors, and the difference and sum of the two
detectors give access to the variances of the generalized
quadratures $X_-$ and $P_+$ from section \ref{quantcorr}, which are
needed to verify the inseparability criterion.

In our case, the signal fields are non-degenerate and require two
phase-locked local oscillators (LOs) with frequencies separated by 6
GHz. To produce the signals and LOs we seed two 4WM processes inside
the same $^{85}$Rb vapor cell. In the first 4WM interaction we
inject only the pump into the cell and seed the probe and conjugate
ports with the vacuum. In the second we inject the pump and also a
small seed (100 $\mu$W) on the probe port. We use this second 4WM
interaction to produce the local oscillators which are used in the
homodyne detectors that measure the field quadratures output by the
first (unseeded) 4WM process. This method of generating the LOs
eliminates problems that might otherwise limit mode matching. For
instance, a cross-Kerr modulation causes a lensing effect inside the
cell that is different for the probe and conjugate and depends on
the pump intensity. Any change in the parameters would require a
change in the mode matching optics for the LOs if they were not
derived from the same 4WM process. Having the LOs undergo the same
Kerr-lensing as the signals eliminates this problem. Further, since
the LOs are derived from a 4WM process that is driven by the same
pump laser as for the signal fields, their phase difference with the
signals is stable, which eliminates the need to actively stabilize
the LO phases.

The signal fields themselves consist of two-mode squeezed vacuum
states, where ``two-mode'' entails the two separate output fields,
one at the probe frequency and the other at the conjugate frequency.
The generalized quadratures for the two-mode squeezed vacuum output
fields should exhibit the noise properties discussed in section
\ref{quantcorr}, which outlines noise properties of the amplifier
for either coherent state or vacuum input. The advantages of using
two-mode squeezed vacuum as the signal are apparent in the
requirement that the local oscillator power be much larger than the
analyzed field in homodyne detection. Our LOs consist of 500 $\mu$W
typically. Further advantages will be evident in later sections when
multi-spatial-mode entanglement is discussed.

In order to align the homodyne detectors, the the 4WM process for
the signals is first seeded with a small input ($\sim 100 \mu$W) so
that the output fields are comprised of bright beams. This provides
a visual guide for alignment on the beamsplitters. After optimizing
the interference between the LOs and signals on the beamsplitters,
the seed is blocked so that the signal fields consist of
two-mode-squeezed vacuum. This alignment procedure insures that the
vacuum modes that are analyzed in the homodyne detectors are the
correct spatial modes to analyze in order to observe entanglement.
In other words, overlapping the LOs with a bright two-mode-squeezed
state output from the cell beforehand insures that the the vacuum
modes analyzed in the homodyne detectors also belong to the same
two-mode-squeezed vacuum state.

\begin{figure}[h!]
    \begin{center}
    \includegraphics[width=4in]{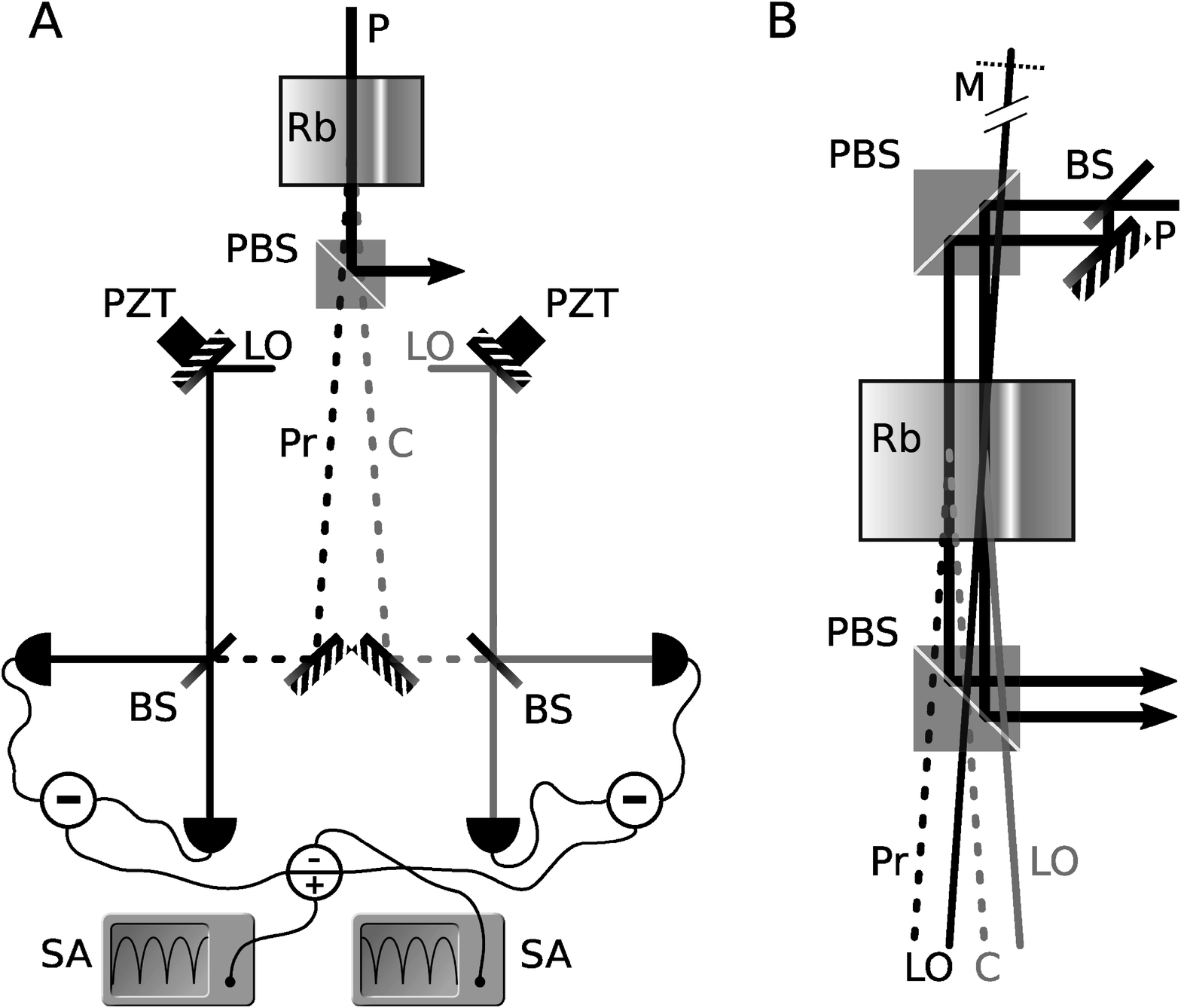}
    \caption{A: Setup of the homodyne detection. P: pump, Pr: probe,
C: conjugate, LO: local oscillator, BS: 50/50 beamsplitter, PBS:
polarizing beamsplitter, PZT: piezoelectric actuator, Rb: rubidium
vapor cell, SA: spectrum analyzer. B: Generation of the LOs used in
the homodyne detectors. Reproduced from ref.~\citenum{science}.
     \label{fig:LOsetup}}
    \end{center}
\end{figure}

Figure \ref{fig:LOsetup}b shows the generation of local oscillator
fields and vacuum squeezed fields. Figure \ref{fig:LOsetup}a shows
the homodyne detection scheme. One of the mirrors in each of the
local oscillator beam paths is mounted on a piezoelectric device.
The piezoelectric actuators are scanned synchronously so that the
relative phases between the local oscillators and the signal fields
are scanned in unison, insuring that the homodyne detectors measure
the same quadratures for the two signal fields at any given time.
The homodyne detector signal as a function of local oscillator phase
gives a generalized quadrature:
$A_\theta=X\cos{\theta}+P\sin{\theta}$. A hybrid junction outputs
the sum and difference of the two homodyne detector signals, which
are recorded simultaneously on two spectrum analyzers. At a local
oscillator phase of zero radians, the difference signal between the
homodyne detectors measures $X_-$, while at a phase of $\pi/2$ the
sum of the signals measures $P_+$.

\begin{figure}[h!]
    \begin{center}
    \includegraphics[width=4in]{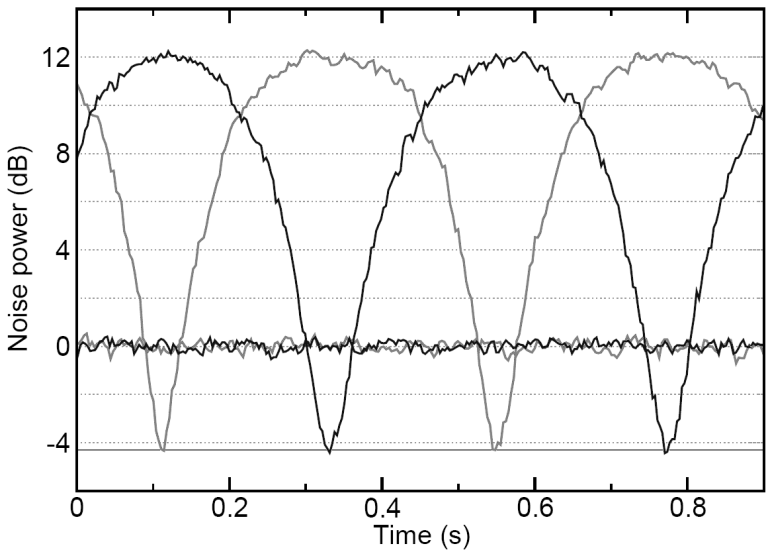}
    \caption{Noise power for the generalized quadratures as a function of local oscillator phase scan. The dark line is the difference between the
    homodyne detectors and its minima represent the $X_-$ quadrature. The gray line shows the summed signal and its minima represent the $P_+$ quadrature.
     Reproduced from ref.~\citenum{science}.
     \label{fig:quad_sq_gauss}}
    \end{center}
\end{figure}

As the LO phases are scanned, the measured quadratures show
correlations or anti correlations in their sum and difference.
Figure \ref{fig:quad_sq_gauss} shows the sum and difference of the
detector signals as a function of spectrum analyzer sweep time,
which amounts to the local oscillator phase. At one point the noise
power of the difference signal is minimized, while the noise of the
summed signal is maximum. At a later time the signals are reversed.
The two noise powers switch places at equal intervals as a function
of local oscillator phase: when one signal is minimized, the other
is always maximized. This indicates that the two homodyne detectors
are measuring the same quadrature for both signal fields, and that
the generalized quadratures being measured at the minima of each
curve are equivalent to the $X_-$ and $P_+$ of section
\ref{quantcorr}.

Figure \ref{fig:quad_sq_gauss} shows noise reduction of -4.3 $\pm$
0.2 dB\footnote{All uncertainties quoted in this paper represent one
standard deviation, combined statistical and systematic
uncertainties.} in each quadrature\cite{science}. This translates to
an inseparability value of $I=0.74 \pm 0.02$, which fulfills the
inseparability criterion $I<2$. The overall detection efficiency was
$90\pm3\%$. The seed for the LOs used for these measurements were
simple gaussian beam profiles. The next section discusses exploiting
the multi-mode nature of the amplifier to produce multi-spatial mode
entangled beams.

\section{Multi-Spatial-Mode Amplification} \label{mm}
Up to now most of the discussion has not made a distinction between the idea of single-spatial-mode
or multi-spatial-mode amplifiers.
In fact, the common thread between each of the systems mentioned at the beginning of section \ref{entanglement}
is that they involve single-spatial-mode operation.
In other words, the signal fields each occupy a single mode of the electromagnetic field, being generated
either inside an optical resonator cavity or a single-mode optical fiber.
Thus, the entanglement generated by these systems is generally between two distinct
modes. In many cases, this is a desirable effect. For instance, a single spatial mode is more easily mode-matched
back into another cavity or optical fiber than a multi-spatial-mode field would be. This inherent strength also
makes these systems difficult to use for quantum imaging, however, where multi-spatial-mode amplification is needed.

The 4WM amplifier we present here does not use a cavity to enhance
the nonlinear gain, and aside from temperature control otherwise
consists of only a simple vapor cell. Since no mode-matching is
required in order to couple the pump and probe into the amplifier,
the spatial modes of the amplifier are determined by the
spatial-phase-matching bandwidth of the gain medium
\cite{jedrkiewicz}. The probe, therefore, can seed several spatial
modes of the amplifier at once. In the special case where the probe
input field is a single mode, the output would still be multi-mode
since the 4WM process would amplify all of the vacuum modes
surrounding the input probe, so that the output state would consist
of many amplified spatial modes regardless of which modes the probe
seeds\cite{prlmultimode}.

One indication of multi-mode behavior is that the amplifier has
appreciable gains for a range of angles between the pump and probe.
In the case of the forward 4WM geometry used here, the angle between
pump and seed is usually a stringent condition \cite{Kumar_Kobolov}.
For our amplifier, this condition is relaxed by two factors. First,
in the absence of dispersion the phase-matching would normally
command all the beams to be collinear, but the appearance of a
strong dispersion of the index of refraction for the probe
\cite{ultraslow} allows the existence of a pair of frequencies for
the pump and the probe such that the 4WM gain is maximum for a
non-zero $\theta$, around $\theta_0 = 7$ mrad. Second, large gain
allows us to use a relatively thin nonlinear medium. This relaxes
the phase-matching condition since the length of the gain medium is
shorter than the coherence length of the nonlinear interaction. This
effectively quasi-phase-matches a range of angles (the shorter the
cell, the larger the range of quasi phase-matched angles). All
spatial modes within the angular bandwidth can be amplified
simultaneously, and quantum correlations can be generated on many
spatial mode pairs at the same time.

\begin{figure}[ht]
    \begin{center}
    \includegraphics[width=4in]{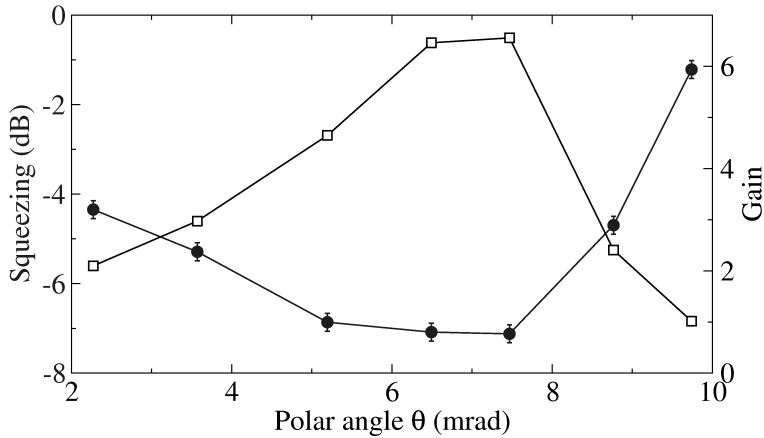}
    \caption{Gain ({\tiny $\Box$}) and intensity-difference noise reduction ($\bullet$) as a function of angle $\theta$.
    Reproduced from ref.~\citenum{prlmultimode}.
     \label{fig:acceptance}}
    \end{center}
\end{figure}

Figure \ref{fig:acceptance} shows the gain and intensity-difference
squeezing as a function of angle between the pump and probe. The
width of the squeezing curve gives an approximate angular acceptance
bandwidth of $\Delta\theta\approx 8$ mrad. Essentially, any input
spatial modes that fall within this angular acceptance bandwidth
will be amplified.

A way to verify the size of an amplified mode in the far field, or
determine the ``resolution'' of the amplifier, is to seed the
amplifier with a probe whose waist is much larger than the pump
inside the amplifying medium. This means that the spread of
k-vectors for the probe is smaller than that of the pump. The
corresponding conjugate will have a diameter much smaller than the
input probe in the far field, limited by the inverse transverse size
of the pump in the gain medium. Under these conditions, the
transverse size of the conjugate is approximately 1 mrad for our
amplifier. The number of spatial modes that the amplifier supports
is then the number of 1 mrad beams that fit into the solid angle of
acceptance. From figure \ref{fig:acceptance}, the acceptance solid
angle is the cone that extends from 2 mrad to 10 mrad, thus the
total number of spatial modes supported by the amplifier is of the
order of $10^2$.

\subsection{Multi-Spatial-Mode Intensity-Difference Noise Reduction}
\begin{figure}[ht]
    \begin{center}
    \includegraphics[width=4in]{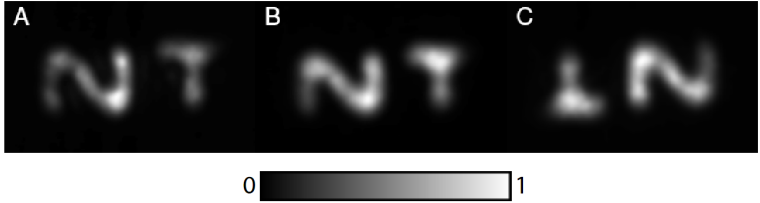}
    \caption{Intensity-difference squeezing in the multi-spatial-mode regime. The images were captured by projecting the beams onto a
    screen in the far field and capturing the images on the screen with a CCD camera. (A) is the image of the probe seed in the far field,
(B) is the image of the output probe in the intermediate field, and
(C) is the image of the conjugate in the far-field. The intensities
are normalized to 1. The different optical conjugation distances for
the probe and the conjugate images show the lensing effect due to
the cross-Kerr interaction with the pump beam, which acts mostly on
the probe. Reproduced from ref.~\citenum{science}.
     \label{fig:NTimage}}
    \end{center}
\end{figure}

A more striking demonstration of the amplifier's multi-mode nature would be an imaging experiment.
Placing an opaque mask in the probe beam path before the
vapor cell changes the shape of the input beam from a Gaussian to that of a spatially-modulated beam (image). As long as the image formed by the mask fits into the spatial bandwidth of the amplifier, the image should be amplified, and a conjugate image should be generated. Figure \ref{fig:NTimage}
shows the input and output of the amplifier for one such mask. Here, the mask cuts the intensity of the probe such that the input
to the amplifier looks like the letters ``N T''.

Measuring the intensity-difference between the probe and conjugate
the same way as for Gaussian beam profiles in earlier sections
showed that the probe and conjugate intensities were correlated with
-5.4 $\pm0.2$ dB of intensity-difference noise
reduction\cite{science}. Thus, the ``N T'' probe and conjugate
images were intensity-difference-squeezed. Next, a spatial filter
(razor blade) was applied to the output images. The T in each image
was blocked so that the detectors measured the intensity of only the
N parts of the image. The intensity-difference between the N parts
was found to be squeezed by -5.2 $\pm 0.2$ dB\cite{science}. When
blocking the N parts and measuring the intensity-difference of the T
parts, the noise reduction was found to be -5.1 $\pm 0.2$
dB\cite{science}. Measuring the difference while blocking the N in
one beam and blocking the T in the other would yield excess noise
\cite{prlmultimode}, which means that the correlations between
images are localized to parts with similar details. Having generated
the ``N T'' image using a single input beam, the presence of
independent correlations between subparts of the output image proves
that the amplifier is multi-spatial-mode
\cite{kobolov1999,Navez2001}. It should be pointed out that the
subparts of the ``N T'' image were independently correlated at
sufficiently high analyzer frequencies ($>3.5$ MHz), but at lower
frequencies cross-talk was present. In other words, below a 3.5 MHz
analyzer frequency, blocking the T would degrade the squeezing
measured between the remaining N subparts. Nonetheless, independent
correlations were observed between subparts for certain analysis
frequencies. Study of the cross-talk effect is beyond the current
scope and will be discussed elsewhere.

The independence of the correlations between subparts of the image
is not unexpected. The noise properties of the amplifier discussed
in earlier sections dealt with single spatial modes of the
electromagnetic field. However, the 4WM amplifier presented here can
be thought of as many single-spatial mode amplifiers working in
parallel. All modes within the spatial bandwidth of the amplifier
are amplified simultaneously, and pairs of spatial modes exhibit the
noise properties discussed in section \ref{quantcorr}. In this
context, the ``N T'' image is clearly a superposition of spatial
amplifier modes. Thus the output images in fact consist of many
subparts that are independently correlated. The size of these
correlated subparts is probably well-approximated by the 1 mrad beam
size given at the beginning of this section.

Lastly, if the multi mode amplifier exhibits intensity-difference
noise reduction, it follows from previous sections that the
amplifier output states also exhibit entanglement for multiple
spatial modes simultaneously. The next section discusses the
generation of quantum-entangled images.

\section{Quantum-Entangled Images}

To detect entanglement between multi-spatial-mode beams, or
image-carrying beams, the experiment of section \ref{entanglement}
was repeated with a mask in the beam path of the seed used to
generate the LOs. The LO shape selects which two-mode-squeezed
vacuum modes the homodyne detectors measures. Changing the angle,
size, or general shape of the LO beam all have an effect on exactly
which spatial modes are analyzed. However, as long as the LO beams
fit within the spatial bandwidth of the amplifier, there will always
be a set of two-mode-squeezed vacuum modes that the bright part of
the LOs can overlap with and analyze in the homodyne detectors. In
general, this means that an arbitrary local oscillator shape can be
used to interrogate the signal fields, as long as the LOs overlap
the correct vacuum modes in the homodyne detectors for both the
probe and conjugate, as discussed in section \ref{entanglement}.

\begin{figure}[h]
    \begin{center}
    \includegraphics[width=4in]{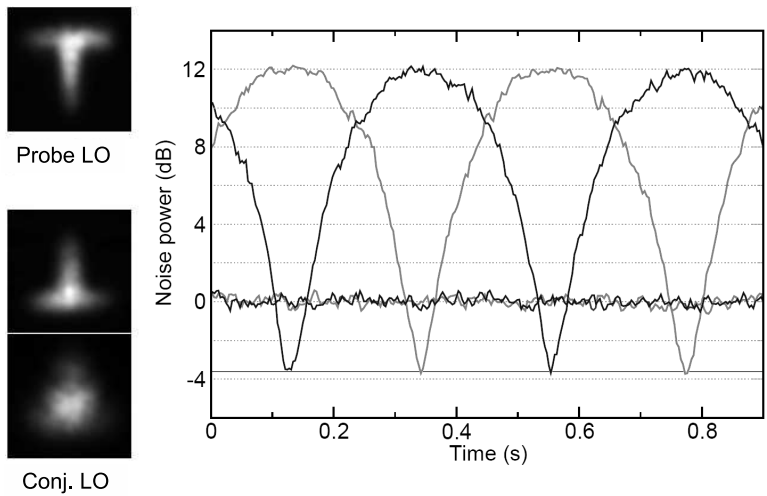}
    \caption{Noise power for the generalized quadratures as a function of local oscillator phase scan. The dark line is the difference between the
    homodyne detectors and its minima represent the $X_-$ quadrature. The gray line shows the summed signal and its minima represent the $P_+$ quadrature.
    The images on the left show the LOs at the homodyne detector. The conjugate image is shown twice, once near the homodyne detector
    where it appears distorted and once in the far field. The difference between the images at the homodyne detectors arises from the fact that
    the probe and conjugate experience different Kerr-lensing effects in the nonlinear medium, and thus have different image planes.
    Reproduced from ref.~\citenum{science}.
     \label{fig:quad_sq_image}}
    \end{center}
\end{figure}

A mask was placed in the seed beam path which imprinted the image of
a letter ``T'' onto the seed and subsequently both LOs. Figure
\ref{fig:quad_sq_image} shows the results of the homodyne detection
measurements for this case, along with the actual LO images at the
homodyne detectors.

Figure \ref{fig:quad_sq_image} shows -3.6 $\pm$ 0.2 dB of noise
reduction in each generalized quadrature\cite{science}. This again
satisfies the inseparability criterion. Therefore, we have shown
that an arbitrarily shaped local oscillator can be used in order to
measure the entanglement in our system. This means that the system
is inherently multi mode, and arbitrary images, within the spatial
bandwidth of the 4WM process, can be amplified as quantum-entangled
images. As a further example we show below in figure
\ref{fig:catmodes} the output images when we use a ``cat face'' mask
on the input. The recorded noise reduction is -1 dB in each
quadrature, showing that the entire images are entangled with one
another between the two-mode-squeezed vacuum signal beams.

\section{Conclusion}
In this paper we have discussed the quantum noise properties of our
phase-insensitive 4WM amplifier system, and we have shown that the
device amplifies and entangles multiple spatial modes
simultaneously. Further, we have shown that our 4WM amplifier can be
used for quantum imaging by demonstrating entanglement between
multiple spatial modes in the form of images. The subparts of the
images, or smaller details, are also independently correlated.
Future work will involve characterizing how the nature of the
entanglement of smaller details changes as the various parameters of
our system change.
\begin{figure}[h]
    \begin{center}
    \includegraphics[width=4in]{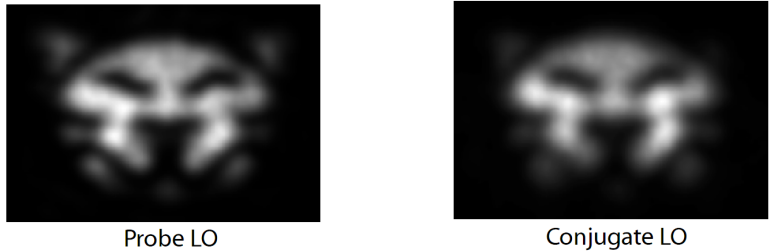}
    \caption{Images of the local oscillators used to measure entanglement in a ``cat face'' mode. The squeezing was -1 dB in each quadrature.
    Reproduced from ref.~\citenum{science}.
     \label{fig:catmodes}}
    \end{center}
\end{figure}

\end{document}